\documentclass[prd,aps,twocolumn]{revtex4}
\usepackage[utf8]{inputenc}

\usepackage{graphicx}
\usepackage{color}
\usepackage{xcolor}

\begin{document}

\title{Model of Cosmic Ray Propagation in the Milky Way at the Knee}
\author{G.~Giacinti$^{1,2}$  and D.~Semikoz$^{3}$}
\affiliation{$^1$Tsung-Dao Lee Institute, Shanghai Jiao Tong University, Shanghai 201210, P. R. China}
\affiliation{$^2$School of Physics and Astronomy, Shanghai Jiao Tong University, Shanghai 200240, P. R. China}
\affiliation{$^3$Astroparticule et Cosmologie, Université de Paris Cite, CNRS,  F-75013 Paris, France}
\begin{abstract}
    We present a new model of anisotropic cosmic ray propagation in the Milky Way, where cosmic rays are injected at discrete transient sources in the disc and propagated in the Galactic magnetic field. In the framework of our model, we show that the cosmic ray spectrum is time-dependent and space-dependent around the energy of the knee. It has a major contribution of one or a few nearby recent sources at any given location in the Galaxy, in particular at the position of the Solar system. We find that the distribution of $\sim$ PeV cosmic rays in our Galaxy is significantly clumpy and inhomogeneous, and therefore substantially different from the smoother distribution of GeV cosmic rays. Our findings have important implications for the calculation and future interpretation of the diffuse Galactic gamma-ray and neutrino fluxes at very high energies.
    %This result is consistent with dramatic change of dipole anisotropy at 100 TeV scale. 
\end{abstract}
\maketitle
\section{Introduction}

A pronounced change in the cosmic ray spectrum, called the \lq\lq cosmic ray knee\rq\rq\/, occurs at the energy $E_{\rm k}\simeq 4$\,PeV. At this energy, its spectral index changes from $\beta\simeq 2.7$ below, to $\beta\simeq 3.1$ above $E_{\rm k}$. This feature was discovered in 1958, but since then, its nature is still the subject of discussions. Several astrophysical explanations have been suggested for the knee. 
They are divided in two main categories: Either the knee is linked to the properties of cosmic ray sources, or it is linked to those of cosmic ray propagation.

Two specific examples of models within the former category are those proposed by Hillas~\cite{Hillas:2005cs} and by Zatsepin et al.~\cite{Zatsepin:2006ci}. These models require at least two  populations of Galactic cosmic rays, one dominating the spectrum below the knee, and the other above. A natural explanation associates these two populations of sources with two different types of supernova (SN) progenitors: Cosmic rays below the knee might be accelerated, e.g., in SN explosions of isolated stars with  masses  $M=8-15M_\odot$, while cosmic-rays with higher energies would be generated by massive OB or Wolf-Rayet stars in superbubbles~\cite{Zatsepin:2006ci}. A characteristic feature of this model is a fast change of the mass composition around $10^{16}$\,eV,

In the second scenario, the knee could be due to a transition from pitch angle scattering to Hall diffusion or drift along the regular magnetic field~\cite{1993A&A...268..726P,Candia:2002we,Candia:2003dk}.
Another possibility is the case suggested in Refs.~\cite{Giacinti:2014xya,Giacinti:2015hva}, where the knee energy corresponds to the rigidity at which the 
cosmic ray Larmor radius, $R_{\rm L}$, is of the order of the correlation length, $L_{\rm c}$, of the turbulent magnetic field in the Galactic disc. In both cases, a transition from large-angle to small-angle scattering or Hall diffusion is expected as a result. Therefore, the energy dependence of the confinement time changes, which in turn induces a steepening of the cosmic ray spectrum~\cite{1971CoASP...3..155S,1993A&A...268..726P,Candia:2002we,Candia:2003dk,Giacinti:2014xya,Giacinti:2015hva}.

It is difficult to distinguish between different cosmic ray knee models based on the present cosmic ray data. However, in recent years, new gamma-ray data has been published from individual sources at 100 TeV by Tibet ASgamma \cite{Amenomori:2019rjd}, HAWC \cite{HAWC:2019tcx} and LHAASO \cite{2021Natur.594...33C}. Some of these sources are most likely to be leptonic, such as the Crab~\cite{Amenomori:2019rjd}, but others may be be hadronic. Another important result is the recent detection of the diffuse gamma-ray emission from the Galactic plane as measured by the Tibet ASgamma experiment \cite{TibetASgamma:2021tpz}. The dominant part of this emission does not come from point sources, and it can contain a significant emission from cosmic ray interactions with the interstellar gas. The interpretation of this diffuse emission can be made by combining it together with that measured by Fermi LAT \cite{Koldobskiy:2021cxt}.

Very recently, the LHAASO experiment has measured this emission with much higher statistics~\cite{LHAASO:2023gne}, which even allows to study details in its spectrum. The interpretation of this data has been done recently within the framework of an isotropic cosmic ray diffusion model in Ref.\cite{Zhang:2023ajh}.

Further important information comes from neutrino telescopes. Contrary to very high energy gamma-rays, neutrinos can only come from hadronic interactions. Some evidence of a signal from our Galaxy was found in the cascade data of IceCube \cite{Neronov:2015osa}. Recently, independent evidence was found in the muon track data \cite{Kovalev:2022izi}. In the future, larger multi-km3 neutrino telescopes will be able to test the hadronic nature of the Galactic gamma-ray emission.

In this work, we propose a new model of anisotropic cosmic ray propagation in our Galaxy, where cosmic rays are injected at discrete transient sources in the Galactic disc and propagated individually in a model of the Galactic magnetic field. Our new model can be used to study future gamma-ray and neutrino data in great detail. One of main advantages of our model is that it allows to study, at the same time, point sources, extended sources of different ages, sizes and shapes (incl. asymmetric shapes), as well as the true, global diffuse emission from very old sources.  This cannot be made with existing isotropic cosmic ray diffusion models.

\section{Dynamical model for anisotropic propagation of Cosmic Rays}

Cosmic-ray fluxes at GeV energies, as well as Fermi-LAT gamma-ray data, can be well described by isotropic cosmic ray diffusion models like GALPROP \cite{Vladimirov:2010aq,Porter:2021tlr}.
%or DRAGON2 \cite{Evoli:2016xgn}.
However, the diffusion coefficient required to reproduce the measured B/C ratio is very high: $D_0 \approx 3 \cdot 10^{28}$ ${\rm cm}^2/{\rm s}$ at GeV energies. Indeed, in isotropic turbulence without any large-scale magnetic field, such a value for $D_0$ would correspond to a turbulent 
magnetic field strength of $B \approx 3 \cdot 10^{-11}$\,G \cite{Giacinti:2017dgt}, whereas the observed Galactic magnetic filed strength is several orders of magnitude higher, $B \sim 1\,\mu$G \cite{Jansson:2012pc}. This conceptual problem has to be solved in some way. One obvious solution is to take into account the regular magnetic field of our Galaxy. In this case, the propagation of cosmic rays is anisotropic with essentially two diffusion coefficients, one along the regular field, $D_{||}$, and the other in the perpendicular direction, $D_{\perp}$. If both the regular and turbulent magnetic fields are of same order ($\mu G $ scale), $D_{||}$ will be responsible for the escape of cosmic rays from our Galaxy \cite{Giacinti:2017dgt}. The main difference between isotropic and anisotropic diffusion models is that the number of locally contributing sources at any point in our Galaxy is reduced in the anisotropic case, because of the small value of $D_{\perp}$. This effect should increase the clumpiness and inhomogeneity of the sea of PeV cosmic rays in our Galaxy.

At GeV energies, where all Galactic SNe contribute to the cosmic ray flux, the anisotropic diffusion model is not distinguishable from the isotropic one. The difference in their predictions for local cosmic ray observables lies in the fluctuations of the local cosmic ray flux, which are always small even in the anisotropic case, due to the large number of sources contributing to the flux at GeV energies. The same is true for gamma-ray observations, in which local fluctuations are further smoothed by the integration over the line of sight.

On the contrary, at PeV energies, the situation is quite different. The escape time of cosmic rays from the Galaxy is decreased by, e.g., a factor 100 for Kolmogorov turbulence. Moreover, the number of sources which are able to accelerate cosmic-rays to PeV energies is expected to be dramatically smaller than at low energies, possibly reduced by a factor 10 or even 100. Indeed, state-of-the-art studies of cosmic-ray acceleration at supernova remnant shocks suggest nowadays that only a small fraction of all supernovae should be able to accelerate cosmic-rays up to the knee or beyond, see for example~\cite{2013MNRAS.431..415B,2018MNRAS.479.4470M}. These studies found that only core-collapse supernovae occurring in dense stellar winds may reach or exceed PeV energies, while most supernovae would only reach $\sim 100$\,TeV to a few 100s of TeV, falling short of reaching the energy of the knee by about one order of magnitude. Moreover, gamma-ray observations of supernova remnants also seem to confirm these theoretical predictions, with many remnants (including some of the {\it a priori} most promising ones, such as Cas~A) displaying a turnover of their gamma-ray spectra beyond $\sim 10$\,TeV. Also, the relatively small number of PeVatrons detected by LHAASO~\cite{2021Natur.594...33C} and the relative softness of their very-high-energy gamma-ray spectra seems to be in line with this expectation.

As a result, at PeV energies, the two models are very different. In the isotropic model, the PeV cosmic ray distribution in the Galactic disc is still homogeneous with only small local fluctuations. On the contrary, our anisotropic model predicts large fluctuations at PeV energies, as well as a significant reduction of the global diffuse gamma-ray flux compared to the contribution from individual gamma-ray sources.

In the following, as an example of Galactic magnetic field model, we take the state-of-the-art Jansson-Farrar model (JF12) \cite{Jansson:2012pc,Jansson:2012rt}. The original turbulent magnetic field strength was overestimated in this model, leading to an overestimation of the cosmic ray confinement time in the Galaxy. In order to reproduce the measured B/C ratio, we reduce the value of the turbulent magnetic field strength by a fixed factor everywhere in the Galaxy, according to the results of our previous studies \cite{Giacinti:2014xya,Giacinti:2015hva}. In the following, we use the outer scale $L_{\max}=25$\,pc for the turbulence.

%%%%%%%%%%%%%%%%%%%%%%%%%%%%%%%%%%%%%%%%%%%%%
\begin{figure}
    \includegraphics[width=0.9\linewidth]{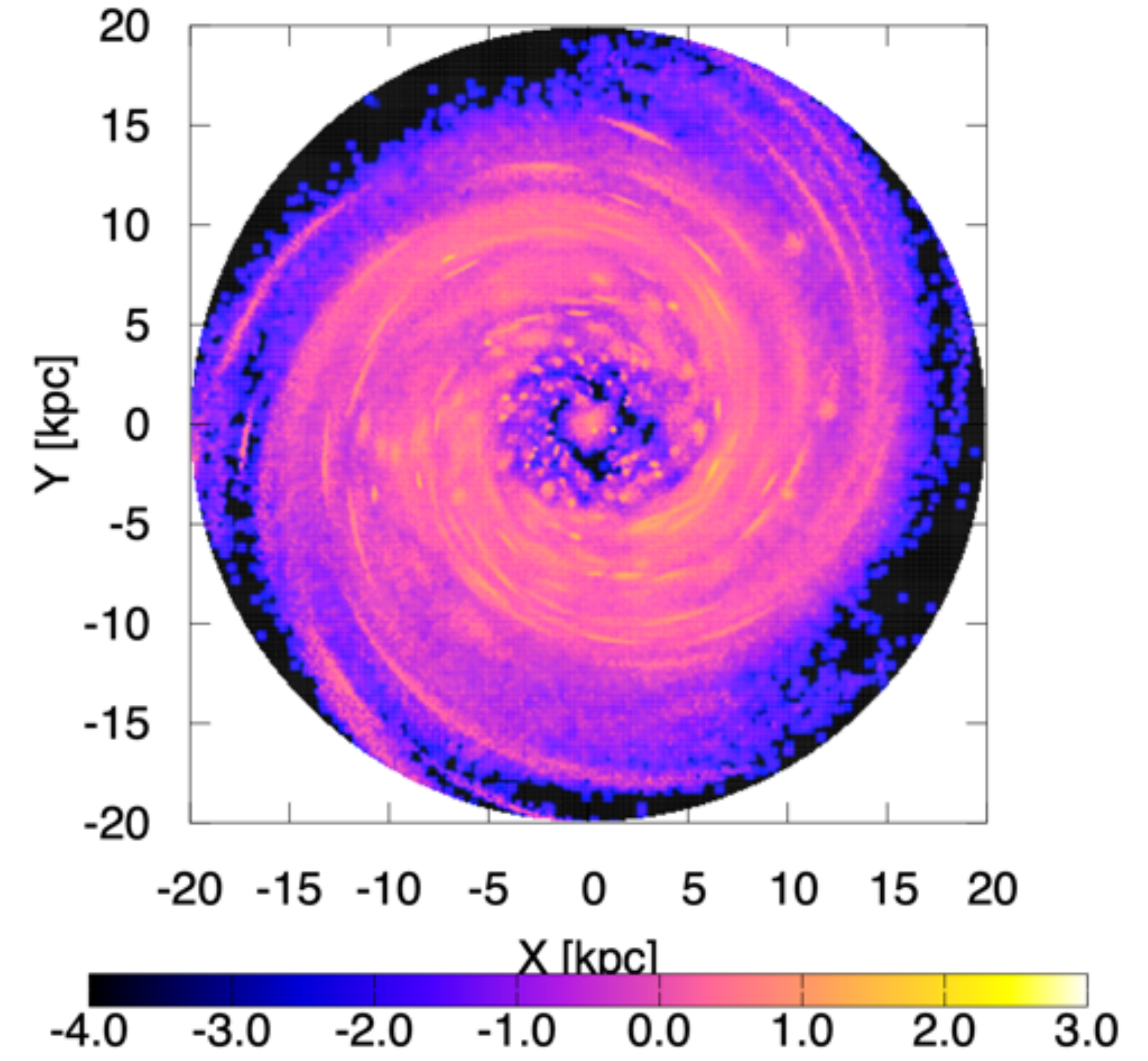}
    \includegraphics[width=0.9\linewidth]{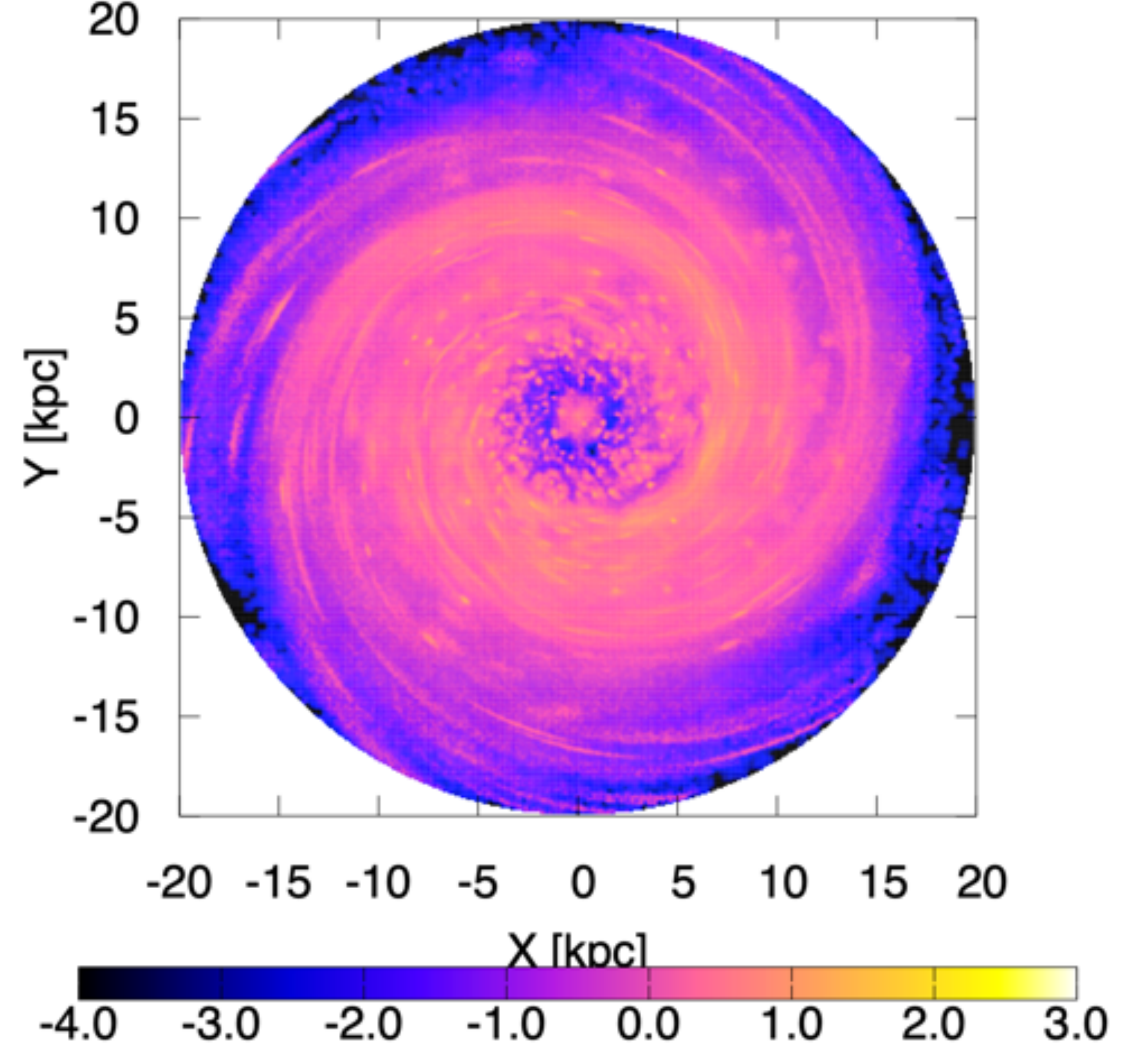}
    \caption{Calculated distributions of $E=1$ PeV cosmic ray protons in the Galactic disc at a given time, as seen from above. In the upper panel, we assume that 1.6\% of all SNe accelerate cosmic rays to PeV, and in the lower panel, we assume that 10\% of SNe accelerate to PeV. The cosmic ray flux is normalised to the value measured at Earth, and the color bar is in log10 scale.}
    \label{fig:map_1PeV}
\end{figure}
%%%%%%%%%%%%%%%%%%%%%%%%%%%%%%%%%%%%%%%%%%%%%

%%%%%%%%%%%%%%%%%%%%%%%%%%%%%%%%%%%%%%%%%%%%%
\begin{figure}
    \includegraphics[width=0.9\linewidth]{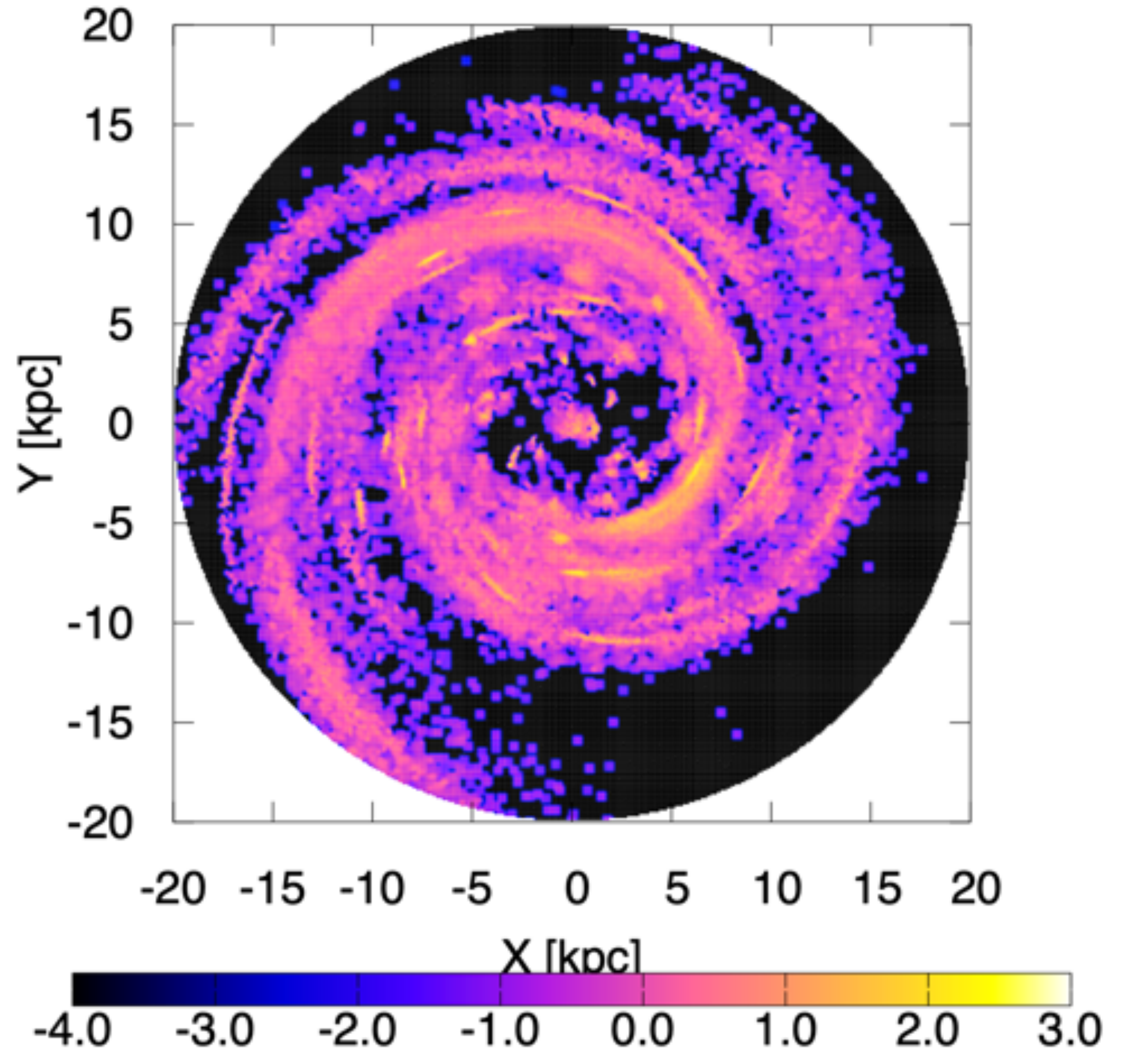}
    \includegraphics[width=0.9\linewidth]{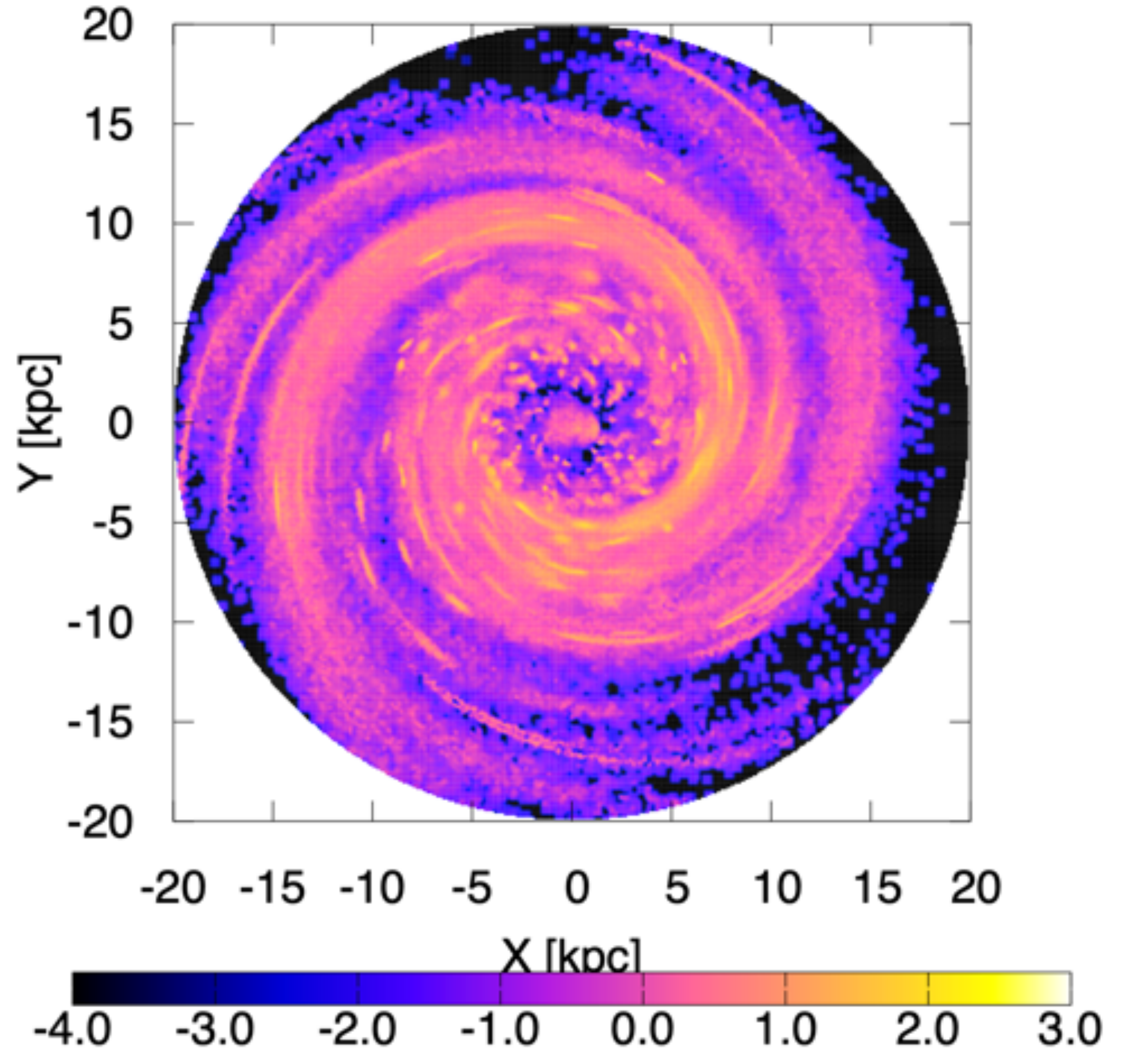}
    \caption{Same as in Fig.~\ref{fig:map_1PeV}, but for cosmic ray protons with $E=10$\,PeV. In the upper panel, we assume that 1.6\%  of all SNe accelerate cosmic rays to 10\,PeV. In the lower panel, we assume that 10\% of SNe accelerate to 10\,PeV.}
    \label{fig:map_10PeV}
\end{figure}
%%%%%%%%%%%%%%%%%%%%%%%%%%%%%%%%%%%%%%%%%%%%%

We define the locations of a few hundreds of \lq\lq reference sources\rq\rq\/ in the Galactic disc at $z=0$, up to a Galactocentric radius $R=18$\,kpc. These fixed reference source locations provide a good sampling of the Galactic disc, such that approximately every reference source within $R=15$\,kpc from the Galactic center is situated at no more than $\approx 1$\,kpc from the nearest other reference source.
We propagate individual cosmic rays in the JF12 model, after injecting them at these designated locations. For each reference source, we calculate $10^3$ cosmic ray trajectories at each of the following five energies: $E=1,3,10,30$, and 100\,PeV. We save the cosmic ray positions in space at fixed, logarithmically-spaced moments in time between 30\,yr and 30\,Myr. This provides us with reusable samples of cosmic ray positions from reference sources, and at given reference times after escape from these sources.

In any of our simulations, we generate randomly the locations and ages of all the SNe that occured within the last 30\,Myr in the Galaxy. We assume that there are on average 3 SNe per century, and that their distribution in the Galaxy follows the formula used in our previous works \cite{Giacinti:2014xya,Giacinti:2015hva}. For each SN, we then search for the nearest \lq\lq reference source\rq\rq\/ in the above list. We then move the locations of its $10^3$ cosmic rays to the actual location of the SN, by doing a small translation of the cosmic ray locations along the Galactocentric radius, and a slight rotation around the Galactic center, such that the location of the reference source is moved to the location of the actual source. We calculate the locations of the cosmic rays according to the actual age of the SN, by interpolating their locations in the two nearest time bins. In this way, we can calculate the cosmic ray flux coming from any source in the Galaxy, at any point in space and time. This allows us to simulate the evolution with time of the cosmic ray distribution in the Galaxy at $\sim (1-100)$\,PeV energies.

In order to calculate the cosmic ray flux at any point in the Galactic disc, we divide it in (100\,pc)$^3$ bins up to a radius of 20\,kpc in the Galactic plane and up to $z=\pm 400$\,pc from the Galactic plane. We calculate the cosmic ray density in each bin assuming that the original cosmic ray flux injected at each source follows a power law $E^{-\alpha}$ up to a maximum energy $E_{max}$, and that the total energy deposited in cosmic rays by each SN is $10^{50}$\,erg between 1\,GeV and $E_{max}$.

We find that, at the Earth position (8.2\,kpc from the Galactic center), 90 \% of the cosmic ray flux at PeV energies comes from sources that are simulated from the 4 nearest \lq\lq reference sources\rq\rq\/ around the Earth. Since we are mostly interested in studying the cosmic ray flux at the Earth here, we simulate, for a greater precision, $10^4$ cosmic ray trajectories from each of these 4 reference sources, instead of the $10^3$ trajectories used for the other reference sources.

In Fig. \ref{fig:map_1PeV}, we show two examples of cosmic ray distributions in the Galactic disc at $E=1$ PeV, averaged over the altitudes $-100$\,pc~$<z<100$\,pc. The upper panel corresponds to the case where only $1.6$\% of all SNe contribute to the cosmic ray flux at this energy, whereas the lower panel corresponds to the case where $10$\% of SNe contribute. In both panels, we plot the relative cosmic ray fluxes, normalized to the value measured at Earth. The color bars are in $log10$ scale, i.e., the value \lq\lq 0\rq\rq\/ corresponds to the flux at the Earth.

In the case with the source density of $1.6$\% of all SNe, we assume that the sources inject cosmic rays with a power law spectrum $\propto E^{-\alpha}$ with $\alpha=2.2$ and $E_{max}>1$ PeV. In this case, the time-averaged value of the cosmic ray flux at Earth at $E=1$ PeV has a value similar to the value measured by the KASCADE experiment. If we use the same injected spectrum for the case with 10\% of SNe, the calculated cosmic ray flux overshoots the KASCADE data by a factor 6 at 1\,PeV. However, this is only 3 times larger than the cosmic ray proton flux measured by IceTop. One way to make our calculated flux consistent with a source density of 10\% of SNe is to take an injected spectrum with $\alpha=2.3$. In this case, the calculated spectrum at PeV energies is between the KASCADE and IceTop data points.

One can clearly see in Fig. \ref{fig:map_1PeV} that our calculated distribution of PeV cosmic rays in the disc is substantially more patchy and inhomogeneous than any of the predictions from standard Galactic cosmic ray propagation models, especially those relying on isotropic cosmic ray diffusion. The case with 10\% of SNe (lower panel) is much smoother in large parts of the Galactic disc than the case with $1.6$\% of SNe (upper panel), but it is still very far from the usually smoother predictions from existing Galactic cosmic ray propagation models.

In Fig. \ref{fig:map_10PeV}, we provide the same calculations at 10\,PeV, above the energy of the knee. One can clearly see that the situation there is even more extreme than at 1\,PeV. The cosmic ray distribution in the Galaxy becomes increasingly more patchy and inhomogeneous. Only a few young SNe contribute to the cosmic ray flux in the Galaxy in the case with $1.6$\% of SNe (upper panel). A bigger number contribute in the case with 10\% of SNe, though still small enough not to give a smooth distribution of cosmic rays. Cosmic ray densities vary by more than four orders of magnitude depending on the observer position and time, even at the same distance from the Galactic center. At these energies, only a few SNe dominate the cosmic ray flux at almost any location of the Galaxy, including at the position of the Earth.

\section{Cosmic ray proton spectrum variability at the knee}

%%%%%%%%%%%%%%%%%%%%%%%%%%%%%%%%%%%%%%%%%%%%%
\begin{figure}
    \includegraphics[width=\linewidth]{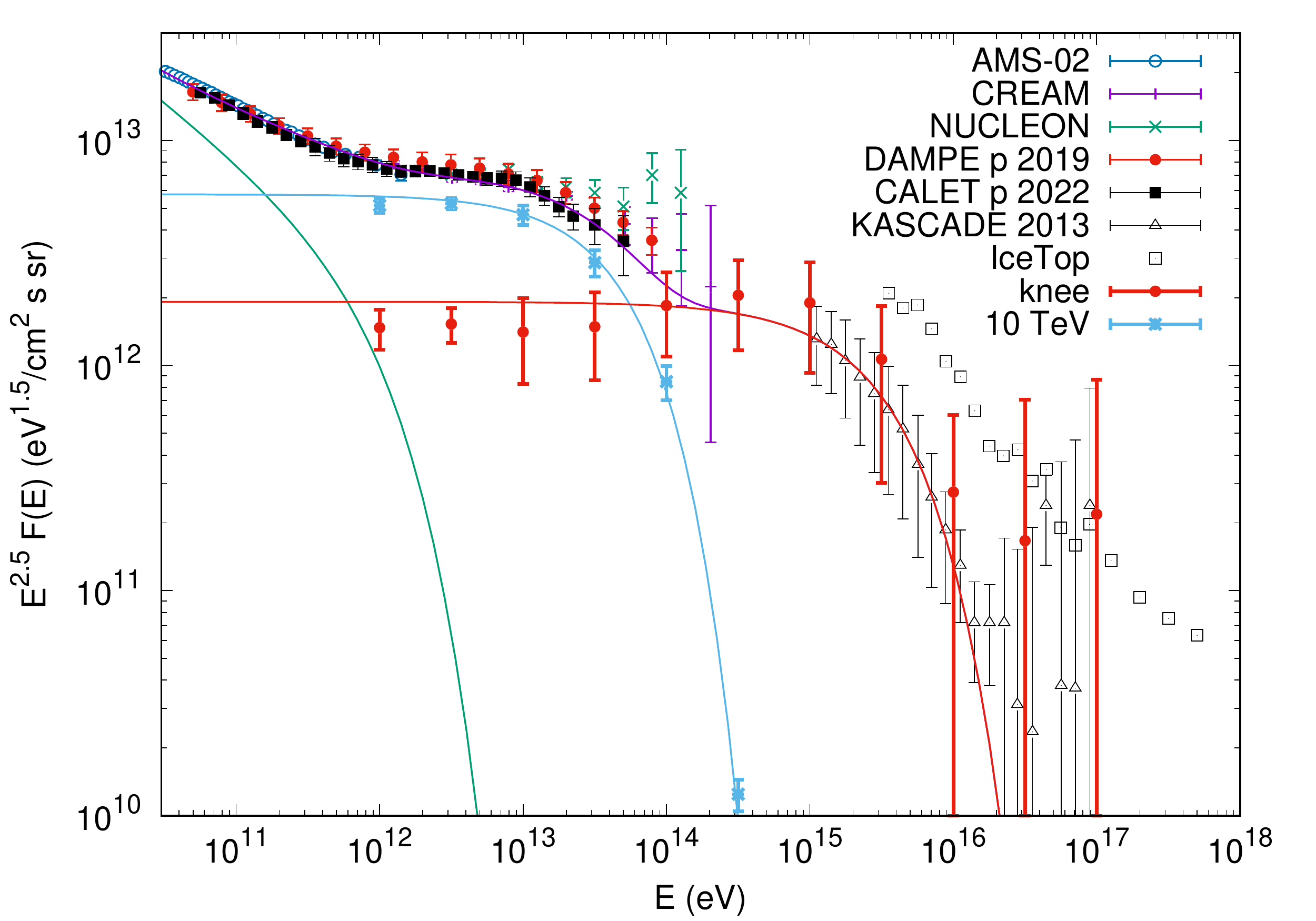}
    \includegraphics[width=\linewidth]{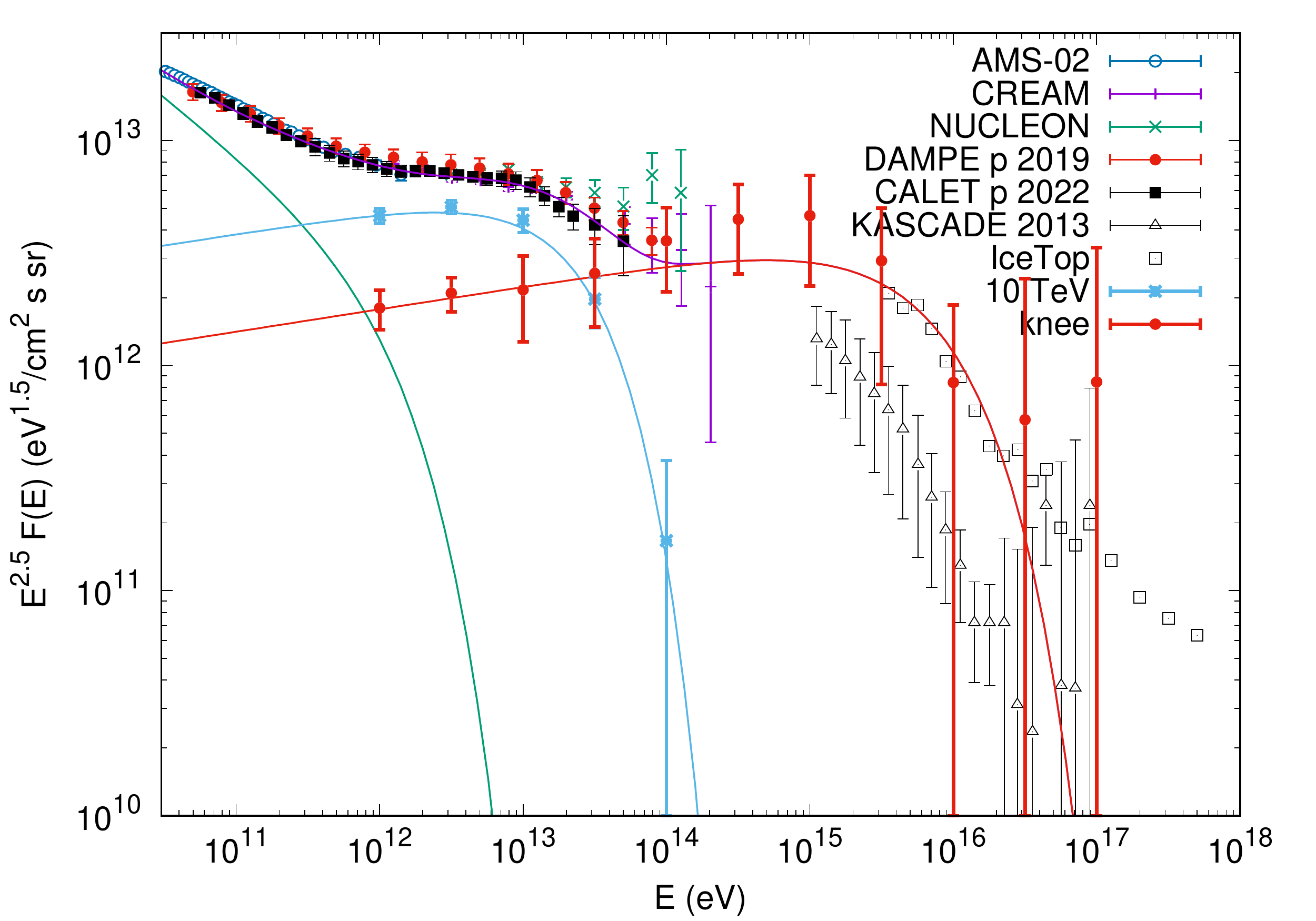}
    \caption{Cosmic ray proton fluxes calculated at the Earth location in our model, for two populations of sources that fit the data from 1\,TeV to 10\,PeV. In the top panel, we inject an $E^{-2.2}$ spectrum at the sources, which fits the KASCADE data at the knee. In the lower panel, we inject an $E^{-2.1}$ spectrum at the sources, which fits the IceTop data at the knee. The errorbars on the modelled fluxes represent their variations with time in our simulations, over Myr timescales. The flux at the knee is dominated by a source population with a density of 1.6\% of all SNe. The flux at 10\,TeV is dominated by the other source population. It is shown with the blue errorbars. We also plot with errorbars the cosmic ray proton fluxes as measured by AMS-02 \cite{AMS:2015tnn}, DAMPE \cite{DAMPE:2019gys}, CREAM \cite{Yoon:2017qjx}, NUCLEON \cite{Gorbunov:2018stf}, CALET \cite{CALET:2022vro}, KASCADE \cite{Apel:2013uni} and IceTop \cite{IceCube:2019hmk} experiments.}
    \label{fig:p_spectrum}
\end{figure}
%%%%%%%%%%%%%%%%%%%%%%%%%%%%%%%%%%%%%%%%%%%%%

%%%%%%%%%%%%%%%%%%%%%%%%%%%%%%%%%%%%%%%%%%%%%
\begin{figure}
    \includegraphics[width=\linewidth]{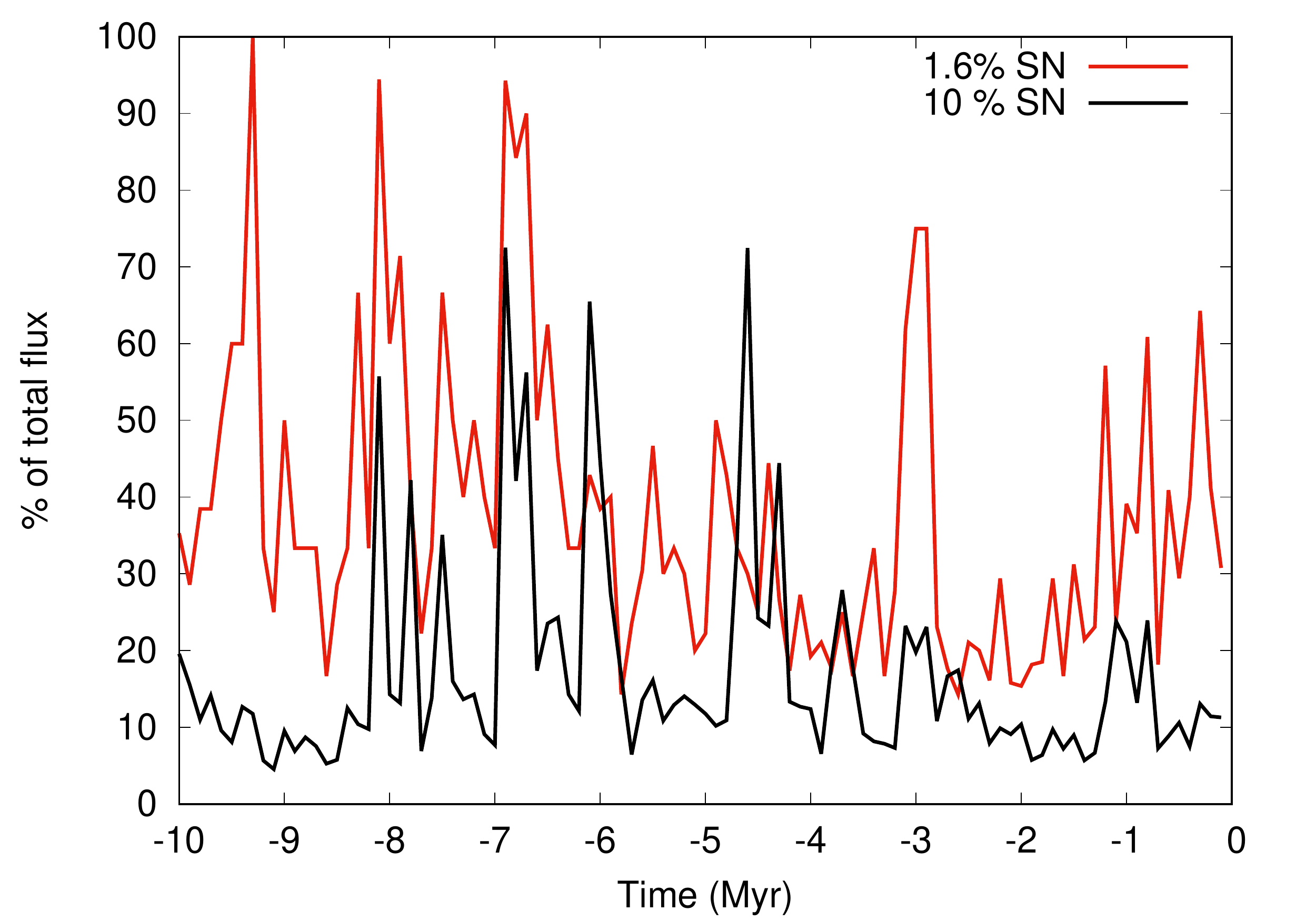}
    \caption{Fraction of the total cosmic ray flux at Earth at $E=1$ PeV that comes from the source with the highest contribution to this flux as a function of time. The red line corresponds to a source density of 1.6\% of all SNe, and the black line corresponds to 10\% of SNe.}
    \label{fig:time_flux}
\end{figure}
%%%%%%%%%%%%%%%%%%%%%%%%%%%%%%%%%%%%%%%%%%%%%

The total cosmic ray flux at the knee has been well measured by a large number of experiments, but the mass composition at those energies still remains unclear. In Fig.~\ref{fig:p_spectrum}, we plot in the form $E^{2.5} F(E)$ two different examples of such measurements: One by KASCADE \cite{Apel:2013uni} experiment, and the other by IceTop \cite{IceCube:2019hmk}. We also plot the cosmic ray proton spectra measured at Earth, in our epoch, by AMS-02 \cite{AMS:2015tnn}, DAMPE \cite{DAMPE:2019gys}, CREAM \cite{Yoon:2017qjx}, NUCLEON \cite{Gorbunov:2018stf} and CALET \cite{CALET:2022vro}.

In this figure, we plot with the wide red errorbars the results for our simulated cosmic ray flux around the knee with the model presented in the previous Section, and with a source density set to 1.6\% of SNe. These errorbars show the amplitude of the time variations of the local cosmic ray flux in our simulations, over a period of 10\,Myr. To guide the eye, we provide fits of these errorbars with solid lines. We find that major fluctuations in this flux typically happen on a $\sim 100$\,kyr time scale. At the Earth location, the cosmic ray flux in the knee region is found to contain detectable contributions from about 100 sources in our Monte Carlo simulations. However, most of the time, the flux at the knee is dominated by only one or a few sources, leading to these very large fluctuations of the flux in time.

In these simulations, we assume that the cosmic rays are injected at the sources with power law spectra $\propto E^{-\alpha}$. In the upper panel, we show the results for simulations with $\alpha=2.2$, and, in the lower panel, we show results for $\alpha=2.1$. Below the energy of the knee, the cosmic ray spectrum after propagation remains a power law $\propto E^{-\beta}$, but with a slope $\beta=\alpha+\delta$, where $\delta=1/3$ for the Kolmogorov turbulent magnetic field used in our simulations. Indeed, the measured slopes of our cosmic ray spectrum just below the knee are compatible with $\beta=2.5$ in the upper panel, and $\beta=2.4$ in the lower panel. Below 1\,PeV, we continue our simulations down to 1\,TeV, using the method discussed in \cite{Kachelriess:2015oua}, which takes advantage of the fact that results on cosmic ray propagation can be rescaled by $\sim E^{1/3}$ for Kolmogorov turbulence.

We note that the cosmic ray spectrum above the $\sim 10$\,TeV \lq\lq bump\rq\rq\/ can be well reproduced with our population of rare sources: Indeed, as can be seen in Fig.~\ref{fig:p_spectrum}, the average cosmic ray flux that we find from this population with a density of only 1.6\% of all SNe fits well the measured cosmic ray spectrum above the bump. This would be in good agreement with the expectation that only a rare subset of SNe, such as SNe occurring in dense winds, could accelerate particles to $\gtrsim 100$\,TeV energies.

%Total number of sources contributed to the flux at Earth in last 10 Myr is around 150 at knee assuming $10^{50}$ erg energy deposited in CR by each source. Most of the time, the flux at the knee is dominated by only a few sources.

Although we mostly focus on $\gtrsim 10$\,TeV cosmic rays in this work, we discuss lower energies below, for completeness. In our model, we assume that there is another, larger population of sources, which is able to accelerate cosmic rays up to $10-100$\,TeV energies, but not to the knee. The flux from this population of sources is shown with the blue errorbars in Fig.~\ref{fig:p_spectrum}. It dominates the cosmic ray flux in the $\sim 1$\,TeV to $\sim (10 - 100)$\,TeV energy range. To guide the eye, we also provide fits of these errorbars with solid lines. Due to the larger number of sources contributing to this low-energy flux, the variations in time of this flux are smaller than those of the other source population. In these simulations, in order to fit the observed spectrum, we need to assume, e.g., that 12.5\% of SNe accelerate to those energies with the same $E^{-\alpha}$ spectrum, if every source releases $5 \cdot 10^{49}$\,ergs in cosmic rays. We note that there is an obvious degeneracy between the energy released in cosmic rays by each source and the required density of these sources to fit the observed cosmic ray spectrum. Therefore, it is entirely possible that all the remaining $(100-1.6)=98.4$\% SNe do accelerate to these energies and contribute to this low-energy flux, provided that each SN releases on average less energy in cosmic rays in this energy band. In that case, the $\sim 10$\,TeV \lq\lq bump\rq\rq\/ would approximately correspond to the maximum energy reached by most SNe in our Galaxy. A third, softer component is added phenomenologically below $\lesssim 1$\,TeV in Fig.~\ref{fig:p_spectrum}, in order to fit the low-energy end of the spectrum. The nature of this third component is both irrelevant for, and beyond, the scope of the current work. We note that cosmic-rays may diffuse in self-generated turbulence at these low energies~\cite{2013JCAP...07..001A}, an effect that our code is not designed to take into account. In practice, this third component might correspond to the population of cosmic rays advected inside the supernova remnants and released at the end of their lives, while the second component between $\sim 1$\,TeV and $\sim (10 - 100)$\,TeV might correspond to the cosmic rays that escape ahead of the shock during the lifetime of the supernova remnant. In that scenario, most SNe in the Galaxy could contribute to these two components. This is reminiscent of the suggestion proposed in Ref.~\cite{2015MNRAS.447.2224B}.

Finally, we note that another free parameter in the above calculations is the maximum energy to which sources can accelerate cosmic rays. It can be degenerate with the power law index $\alpha$ \cite{Kachelriess:2005xh}.

%In knee region we plot results of our simulations for the case of 1.6\% of SN at different times with squares. Today values are shown with violet color. 
%We also show results  2 Myr and 4 Myr ago. One can see that both spectra of today and 2 Myr case are similar. Contrary, 4 Myr case has much higher flux
%well above observed data. 
%This case corresponds to the simulation, when recent nearby SN give significant contribution to local flux.

In Fig. \ref{fig:time_flux}, we plot the fraction of the cosmic ray flux at Earth at $E=1$ PeV which is due to the source with the largest contribution to this flux. In the case of a source density equal to 1.6\% of all SNe (red line), the contribution from the source with the highest flux to the total flux at Earth varies between 20\% and almost 100\%. In this case, the total number of sources having a non-zero contribution to the flux at Earth in our simulations varies around 100 at any given time, although most of them provide only a very small contribution to it. For a source density equal to 10\% of SNe (black line), the contribution from the source with the highest flux can still be up to about 70\% of the total flux, but, most of time, it stays at the level of 10\%. In this case, the total number of sources having a non-zero contribution to the flux at Earth in our simulations varies around 1000.

At $E=10$\,PeV, the contribution of the source with the largest flux varies between 50\% and 100\% in both cases, however in the case of 1.6\% of SNe, it remains close to maximum for a significant fraction of the time.

\section{Discussion and conclusions}

In this paper, we have presented a new model of anisotropic cosmic ray propagation in our Galaxy around PeV energies, where cosmic rays are injected stochastically at individual transient sources in the Galactic disc and propagated individually in Galactic magnetic field models. With our code, we can simulate the full 3-dimensional distribution of $\sim$ PeV cosmic rays in the Galaxy as a function of time. As a result, we can simultaneously study the contribution from individual sources and the global diffuse cosmic ray flux, as well as estimate its local time and space fluctuations.

The widely-used isotropic cosmic ray diffusion models require a diffusion coefficient that is two orders of magnitude larger than those expected for cosmic rays diffusing in pure isotropic turbulence with $\sim \mu$G strengths \cite{Giacinti:2017dgt}. Anisotropic cosmic ray diffusion models, such as ours, allow to remove this major tension by making cosmic rays escape along the regular Galactic magnetic field direction in the halo. We used here the state-of-the-art Jansson-Farrar Galactic magnetic field model \cite{Jansson:2012pc,Jansson:2012rt}, with a reduced turbulent component. This allows our model to satisfy the constraints on the cosmic ray confinement time from the B/C ratio \cite{Giacinti:2014xya,Giacinti:2015hva} with standard $\sim \mu$G Galactic magnetic field strengths. However, this leads to major changes in the distribution of PeV cosmic rays in our Galaxy.

In particular, we find that our model has two important consequences. First, as illustrated in Figs. \ref{fig:map_1PeV} and \ref{fig:map_10PeV}, the cosmic ray flux in the Galaxy is highly inhomogeneous in space, at PeV energies and above. For practical applications, this means that the standard assumption in gamma-ray astronomy that the line-of-sight emission should be proportional to the summed-up gas density along the line-of-sight does not hold anymore at such energies. Second, as shown in Fig. \ref{fig:time_flux}, the cosmic ray flux at PeV energies can contain a dominant contribution from one single source, due to the small number of sources contributing to the flux at any point in the Galaxy, and in particular at the position of the Earth. At 10 PeV energies, this is almost always guaranteed due to the even smaller number of sources contributing to the flux.

Thus, we expect the cosmic ray distribution in our Galaxy to be strongly inhomogeneous at $\sim$ PeV energies and beyond, and we expect that the local high-energy cosmic ray flux at any point in the Galactic disc displays strong fluctuations with time.

We also expect our findings to have important consequences for the predicted secondary $\gtrsim 100$\,TeV gamma-ray and neutrino emissions from our Galaxy. These emissions should therefore be substantially more patchy and clumpy in our model than in the usual predictions from isotropic cosmic ray propagation models.

\bibliography{knee.bib}

\end{document}